\documentclass[aps,amsfonts,reprint,tightenlines,amssymb,superscriptaddress,twocolumn,floatfix]{revtex4-2}  

\usepackage{graphicx}
\usepackage{dcolumn}
\usepackage{bm}
\usepackage{mathtools}
\usepackage{amsmath}
\usepackage{amsthm}
\usepackage[utf8]{inputenc}

\usepackage{qcircuit}
\usepackage{url}
\usepackage[caption=false]{subfig}

\usepackage{algorithm}
\usepackage[noend]{algpseudocode}
\algrenewcommand\algorithmicdo{}

\makeatletter
\renewcommand{\ALG@name}{Procedure}
\makeatother

\newcommand{\LiPS}{Li$_3$PS$_4$}
\newcommand{\Sring}{S$_8$}
\newcommand{\fieldunit}{V\,\AA$^{-1}$}

\usepackage[linktocpage=true,
  colorlinks=true, 
  pdfborder={0 0 0},
  linkcolor=blue,
  citecolor=red,
  filecolor=yellow,
  urlcolor=blue,
  bookmarks,
  pdfauthor={},
]{hyperref}
\usepackage{orcidlink}

\usepackage{ifthen}
\newcounter{is_qcircuit_used}
\newcommand{\bs}{\boldsymbol}

\begin{document}

\preprint{APS/123-QED}

\title{
Orbital-Free DFT-Assisted Machine-Learned Molecular Dynamics for Electric-Field-Driven Ionic Transport
}

\author{Yusuke Nishiya\orcidlink{0000-0001-6526-0936}}
\email{ynishiya@quemix.com}
\affiliation{
  Department of Physics, 
  Graduate School of Science, 
  The University of Tokyo,
  7-3-1, Hongo, Bunkyo-ku, Tokyo 113-0033, Japan
}
\affiliation{
Quemix Inc.,
Taiyo Life Nihombashi Building,
2-11-2,
Nihombashi Chuo-ku, 
Tokyo 103-0027,
Japan
}

\author{Hiroya Nakata}
\affiliation{
Advanced Technology Research Laboratories, Innovation Center, Idemitsu Kosan Co., Ltd.
1280 Kami-izumi, Sodegaura, Chiba, 299-0293, Japan
}

\author{Yosuke Harada}
\affiliation{
Advanced Technology Research Laboratories, Innovation Center, Idemitsu Kosan Co., Ltd.
1280 Kami-izumi, Sodegaura, Chiba, 299-0293, Japan
}

\author{Yu-ichiro Matsushita\orcidlink{0000-0002-9254-5918}}
\affiliation{
  Department of Physics, 
  Graduate School of Science, 
  The University of Tokyo,
  7-3-1, Hongo, Bunkyo-ku, Tokyo 113-0033, Japan
}
\affiliation{
Quemix Inc.,
Taiyo Life Nihombashi Building,
2-11-2,
Nihombashi Chuo-ku, 
Tokyo 103-0027,
Japan
}
\affiliation{
Quantum Material and Applications Research Center,
National Institutes for Quantum Science and Technology (QST),
2-12-1, Ookayama, Meguro-ku, Tokyo 152-8552, Japan
}

\date{\today}


\begin{abstract}
We propose an orbital-free density functional theory (OFDFT)-assisted machine-learned molecular dynamics method in which field-independent interatomic forces are evaluated using a machine-learned potential and atomic charges that depend on the local environment are obtained from OFDFT calculations. The atomic charges obtained by Bader partitioning of the OFDFT electron density are multiplied by the electric-field vector and added to the forces from the machine-learned potential. 
This approach enables molecular dynamics simulations under an electric field at a lower computational cost than Kohn–Sham DFT (KSDFT)-based molecular dynamics.

As a proof of concept, the method was applied to $\beta$--Li$_3$PS$_4$ bulk and an S$_8$/Li$_3$PS$_4$ heterostructure under periodic boundary conditions. For Li$_3$PS$_4$ bulk, OFDFT yielded Li charges and a total charge of the PS$_4$ unit consistent with those obtained using KSDFT. In the heterostructure, Li ions migrated from the Li$_3$PS$_4$ region into the S-rich region within a simulation time of approximately 170~ps. 
Accompanying this migration, the mean Bader charge of the S atoms that initially formed S$_8$ rings changed from nearly neutral to negative. A comparison with KSDFT for a representative interfacial structure also confirmed that the negative charging of the S atoms upon the arrival of Li and the magnitude of the Li charges were qualitatively reproduced. 
These results demonstrate the possibility of treating field-driven ionic transport and changes in interfacial charge states using a universal machine-learned potential without training an additional material-specific charge-prediction model.
\end{abstract}

\maketitle 

\section{Introduction}
Understanding ionic transport, structural changes, and interfacial reactions under external electric fields at the atomic scale is important for the design of a wide range of ionic materials, including solid electrolytes, electrode materials, and electrochemical interfaces.
Molecular dynamics (MD) is a powerful method for analyzing these phenomena; however, treating spatial and temporal scales close to those of real materials requires both interatomic interactions and responses to external fields to be evaluated accurately and at low computational cost.

In recent years, the emergence of universal machine-learned interatomic potentials has made it possible to rapidly evaluate the energies and interatomic forces of diverse systems without constructing a potential from scratch for each material~\cite{Batatia2025,Rhodes2025}.
However, since electron densities and atomic charges that vary with the local environment are generally not provided as outputs, the force due to an electric field on each atom cannot be evaluated directly.
Formal charges or fixed charges obtained from an initial structure can also be used, but this approximation cannot describe changes in atomic charges accompanying ion migration or changes in the coordination environment.
This limitation becomes particularly serious in systems where the local atomic environment changes substantially, such as electrode--electrolyte interfaces.

In a periodic system, within the linear-response regime, the force arising from a uniform electric field on atom $i$ can be expressed using the Born effective charge tensor as
\begin{equation}
  F^{\mathrm{field}}_{i\alpha}
  =
  e\sum_{\beta} Z^{*}_{i,\alpha\beta}E_{\beta},
  \label{eq:born-force}
\end{equation}
where $e$ is the elementary charge, $Z^{*}_{i,\alpha\beta}$ is the Born effective charge tensor, and $E_{\beta}$ is the $\beta$ component of the electric-field vector.
Shimizu et al.\ constructed a neural network that predicts each component of the Born effective charge from an atomic structure and, by combining it with a neural-network potential, performed MD simulations of crystalline and amorphous Li$_3$PO$_4$ under electric fields~\cite{Shimizu2023}.
In that study, however, the training data for Li$_3$PO$_4$ were generated from density-functional perturbation theory calculations for approximately 18,000 structures.
Application to different compositions or interfaces requires the new collection of Born effective charge data covering the target chemical systems and the retraining and validation of the prediction model.

Another option for treating charge transfer and chemical reactions at classical computational cost is a reactive force field incorporating a charge-equilibration method.
However, ReaxFF parameters strongly depend on the combination of target elements and the reaction environment, and parameters from different development families cannot be combined arbitrarily.
Transferring existing parameters to a new chemical system requires extensive refitting and validation~\cite{Senftle2016}.
For example, ReaxFF parameters have been developed for lithiated sulfur, including Li$_2$S and various Li$_x$S compositions~\cite{Islam2015}, but their reliability is not necessarily guaranteed for electrode--electrolyte interfaces such as \Sring/\LiPS.

It is of course possible to obtain the electron density and evaluate atomic charges according to the local environment by performing Kohn--Sham density functional theory (KSDFT)~\cite{KohnSham1965} during an MD simulation or by performing first-principles MD from the outset.
However, because the computational cost associated with solving and orthogonalizing Kohn--Sham orbitals increases rapidly with system size, long-time MD simulations of interfaces containing thousands of atoms remain difficult~\cite{Mi_review2023}.

By contrast, orbital-free DFT (OFDFT), a type of DFT in which the kinetic energy functional is explicitly given as a functional of the electron density, can be performed at a computational cost of $O(N)$ to $O(N \log N)$ for $N$ atoms. Thus, the electron density of a large-scale system can be evaluated at lower cost than with KSDFT~\cite{delRioCarter_review2018, Mi_review2023,Shao2021}.
This approach may make it possible to rapidly obtain charge information that reflects changes in the local atomic environment without constructing a material-specific charge-prediction model in advance.
On the other hand, the applicability of OFDFT itself to accurate interatomic-force calculations remains limited although structural relaxations and first-principles molecular dynamics driven by OFDFT forces have reproduced KSDFT results for the radial distribution functions of simple liquid metals and for bulk properties such as equilibrium volumes and moduli~\cite{Gonzalez2002, HuangCarter2010PRB, Imoto2021PRR}. For example, even computationally expensive nonlocal kinetic-energy density functionals can predict minimum-energy atomic structures at semiconductor interfaces that differ from those obtained using KSDFT~\cite{Shao_revHC2021}.

In this study, therefore, the field-independent interatomic forces and the forces caused by the electric field are assigned to separate computational methods.
The former are evaluated using a universal machine-learned interatomic potential, whereas the latter are obtained by performing OFDFT calculations and Bader-partitioning the resulting electron density at fixed intervals along the MD trajectory.
The electric-field vector multiplied by the Bader charge of each atom is added to the interatomic force from the machine-learned potential.
This construction extends a universal machine-learned potential to MD simulations under electric fields.

As an example application of this method, we consider a cathode--electrolyte interface consisting of \Sring/\LiPS. We first compare OFDFT charges in bulk \LiPS\ with KSDFT charges and then apply the method to an \Sring/\LiPS\ heterostructure under periodic boundary conditions to examine electric-field-dependent Li-ion migration and the accompanying changes in the charges of interfacial S atoms.

\section{Methods}
\subsection{Hybrid force model}
The force used to update the motion of atom $i$ is defined as
\begin{equation}
  \bm{F}^{\mathrm{tot}}_i(t)
  =
  \bm{F}^{\mathrm{ML}}_i\!\left(\{\bm{R}(t)\}\right)
  +
  q_i(t_k)e\bm{E},
  \quad
  t_k \leq t < t_{k+1},
  \label{eq:total-force}
\end{equation}
where $\bm{F}^{\mathrm{ML}}_i$ is the force from the machine-learned potential without an external field, and $q_i$ is the dimensionless charge in units of the elementary charge $e$. In this study, $q_i$ is obtained by performing Bader charge analysis on the electron density calculated using OFDFT.
The OFDFT calculation and Bader analysis are called at the discrete times $t_k$, and in the present simulations the charges are held constant between two successive electron-density calculations.

The machine-learned interatomic potential represents interatomic forces at zero electric field. Either a universal machine-learned potential or a potential developed specifically by a user for an individual material may be used. In the present model, as represented by Eq.~\ref{eq:total-force}, the calculation of correction forces in directions perpendicular to the electric field, that is, off-diagonal contributions, and forces arising from polarization induced by the applied electric field are outside the scope of the model.
In this study, the interatomic forces were evaluated using Orb-v3~\cite{Rhodes2025} through the Atomic Simulation Environment (ASE)~\cite{Larsen2017}.
The temperature was controlled using the Nos\'e--Hoover dynamics~\cite{Nose1984,Hoover1985} implemented in ASE.


\subsection{Bader analysis using the OFDFT density and a Gaussian core reference density}
DFT~\cite{HohenbergKohn1964} states that the total energy of an electron system can be expressed solely as a functional of the electron density $\rho(\bs{r})$. The electron density that minimizes this functional corresponds to the ground-state electron density. The total energy functional is constructed as follows:
\begin{gather}
E[\rho] = T_S[\rho] + \int v_\text{ext}(\boldsymbol{r})\rho(\bs{r}) d\boldsymbol{r} + E_\text{H}[\rho] + E_\text{XC}[\rho],
\label{eq:etot}
\end{gather}
where $T_S$ is the kinetic energy of the non-interacting system, $E_\mathrm{H}$ is the Hartree energy representing the classical electron--electron interaction, and $E_\mathrm{XC}$ is the exchange--correlation energy.
In OFDFT, the kinetic energy functional is directly given as a functional of the density.
In this study, we assume the use of local pseudopotentials, such as bulk-derived local pseudopotentials (BLPSs)~\cite{Huang2008, HuangCarter2010PRB, Zhou2004PRB, delRio2017}, optimized effective pseudopotentials (OEPPs)~\cite{OEPP2016}, and {\it ab initio} local ionic pseudopotentials (LIPSs)~\cite{Imoto2021PRR}, as the potential $v_\mathrm{ext}$ representing the interaction between the valence electrons and the nuclei with core electrons. Because OFDFT does not treat orbitals, the nonlocal pseudopotentials ordinarily used in KSDFT cannot be used without modification.

The Bader charge assigned to each atom is calculated from the resulting electron density $\rho(\bm r)$ based on the methods of Refs.~\cite{Tang_2009, Sanville_2007, Henkelman2006}.
The valence electron density obtained as the solution of a DFT calculation using pseudopotentials generally does not have a maximum at the atomic positions. Therefore, when this density is directly Bader-partitioned, all valence electrons surrounding atom $A$ may be assigned to neighboring atoms, and both the Bader volume and Bader charge of atom $A$ may become zero.
We therefore define a reference density in which the core-electron density contribution is approximated by Gaussians:
\begin{equation}
  \rho_{\mathrm{ref}}(\bm r)
  =
  \rho_{\mathrm{val}}(\bm r)
  +
  \sum_A
  \frac{Q_A^{\mathrm{ref}}}
       {(2\pi\sigma_A^2)^{3/2}}
  \exp\!\left[
    -\frac{|\bm r-\bm R_A|^2}{2\sigma_A^2}
  \right].
  \label{eq:reference-density}
\end{equation}
Here, $Q_A^{\mathrm{ref}}$ and $\sigma_A$ are the reference amplitude and width, respectively, for each element.
The Bader region $\Omega_A$ is defined as the region enclosed by zero-flux surfaces of the reference density.
Importantly, the Gaussian density is used only to define $\Omega_A$.
The electron count $N_A^\mathrm{val}$ and charge $q_A$ are given by
\begin{equation}
  N_A^{\mathrm{val}}
  =
  \int_{\Omega_A}\rho_{\mathrm{val}}(\bm r)\,d\bm r,
  \qquad
  q_A
  =
  Z_A^{\mathrm{val}}-N_A^{\mathrm{val}},
  \label{eq:bader-charge}
\end{equation}
so that the auxiliary Gaussian is not added to the valence electron count of each atom. Here, $Z_A^\mathrm{val}$ is the number of valence electrons in the isolated atom. In calculations using the projector augmented-wave (PAW) ~\cite{PAW1994}, the all-electron density is used as $\rho_\mathrm{ref}$.
The KSDFT and OFDFT calculations in this study were performed with Quloud based on Quantum ESPRESSO~\cite{Giannozzi2009, Giannozzi2017} and DFTpy~\cite{Shao2021}, respectively.

\section{Results and discussions}
\subsection{Charge evaluation for bulk \LiPS}
To examine the charges obtained from OFDFT, the following four conditions were compared using a 32-atom orthorhombic $\beta$-\LiPS\ cell with lattice constants
$a=6.17\mathrm{\AA}, b=8.02\mathrm{\AA}, c=13.03\mathrm{\AA}$.

Under condition (a), KSDFT calculations were performed using the Perdew--Wang (PW) form~\cite{PerdewWang1992} of the local density approximation (LDA) and the PAW method, whereas under condition (b), the Perdew--Zunger (PZ) form~\cite{PerdewZunger1981} of the LDA and the PAW method were used.
Because the Li $1s$ electrons were treated as valence electrons in these calculations, the cutoff energies for the wave functions and charge density were set to 100~Ry and 400~Ry, respectively.
The resulting charge density was evaluated on an
$80\times100\times160$
real-space grid.

Under condition (c), KSDFT calculations were performed using the PW form of the LDA and Martins--Troullier norm-conserving pseudopotentials (NCPPs).
Only the Li $2s$ orbital was treated as a valence orbital; the cutoff energies for the wave functions and charge density were set to 60~Ry and 240~Ry, respectively, and the real-space grid was set to
$60\times80\times128$ points.
For the above KSDFT calculations, a
$6\times4\times3$
$k$-point mesh was used for Brillouin-zone integration.

Under condition (d), OFDFT calculations were performed using the PZ form of the LDA and the LKT (Luo-Karasiev-Trickey)~\cite{Luo2018} kinetic energy density functional (KEDF).
BLPSs were used as the local pseudopotentials for Li and P, and an OEPP was used for S.
The cutoff energy for the electron density and the real-space grid were set to the same values as those under condition (c). The same electron distributions were used to approximate the core electron densities under conditions (c) and (d). The Gaussian amplitude $Q_A^\mathrm{ref}$ was set to the difference between the atomic number and the number of valence electrons, and the width $\sigma_A$ was set to 0.40~Bohr for all three elements.

The mean Bader charge for each element obtained from each calculation is shown in Table~\ref{tab:lpsbulk_chg}.
A comparison of (a) and (b) shows that, at the same LDA level, the difference in the exchange--correlation functional has almost no effect on the charge of each element. A comparison of (a) and (c) shows that the correction of the core electron density using the present parameters realizes a partitioning that yields similar Bader charges.
The OFDFT results in (d) show that the Li charge is in good agreement with the KSDFT result, whereas the number of electrons around P is overestimated. However, this discrepancy arises from the assignment of charge within the $\mathrm{PS_4}$ framework, and the total charge of $\mathrm{PS_4}$ agrees with the KSDFT result. Therefore, rapid charge evaluation using OFDFT is considered useful, for example, in use cases that focus only on the drift of Li ions along the electric field in the present material and do not regard the modulation of vibrational modes within the $\mathrm{PS_4}$ framework as important.
\begin{table}[ht]
\centering
\caption{Mean Bader charges for each element obtained for the \LiPS crystal. The charges are given in units of the elementary charge $e$.}
\label{tab:lpsbulk_chg}
\begin{tabular}{lccc} \hline
   Calculation condition &Li & P & S \\ \hline
   (a) KSDFT, PAW, LDA-PW & +0.87 & +1.28 & -0.97 \\
   (b) KSDFT, PAW, LDA-PZ & +0.87 & +1.29 & -0.97 \\
   (c) KSDFT, NCPP, LDA-PW & +0.90 & +1.25 & -0.99 \\
   (d) OFDFT, LDA-PZ & +0.87 & +0.06 & -0.66 \\ \hline
\end{tabular}
\end{table}

\subsection{\Sring/\LiPS\ heterostructure}
\begin{figure*}[ht]
    \centering
    \includegraphics[width=0.82 \textwidth]{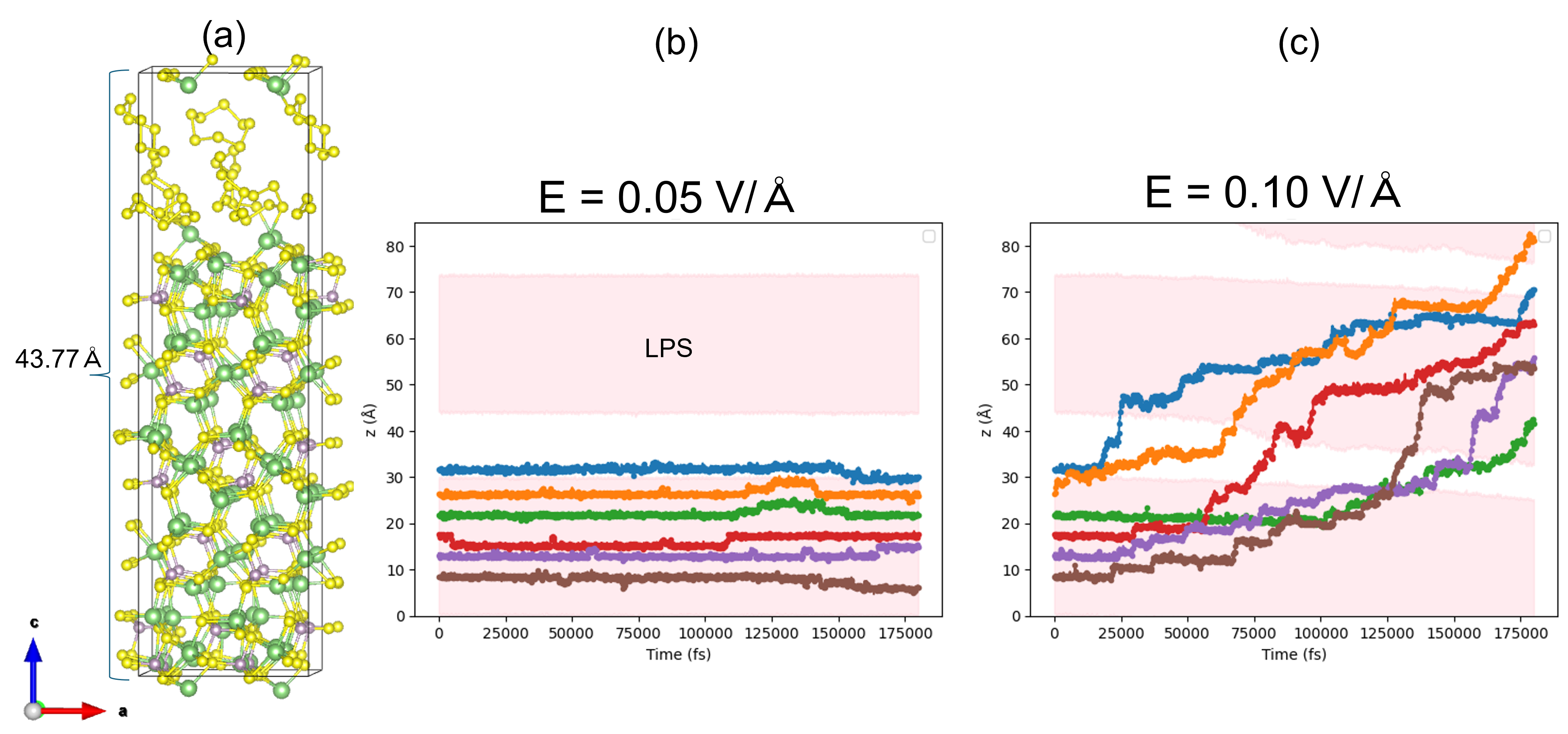}
    \caption{
Application example of the present method to an $\mathrm{S_8/Li_3PS_4}$ heterostructure. (a) Initial structure. The green, gray, and yellow spheres correspond to Li, P, and S, respectively. (b) Trajectories of Li ions at 500~K when an electric field of 0.05~\fieldunit\ was applied along the $c$ axis. (c) Trajectories of Li ions at 500~K when an electric field of 0.10~\fieldunit\ was applied along the $c$ axis. The region containing P atoms was defined as the \LiPS\ region and is shaded pink.
    }
    \label{fig:heterto_short}
\end{figure*}

As a simple model of a cathode--electrolyte interface in a Li battery material, the structure shown in Fig.~\ref{fig:heterto_short}(a) was constructed, and Li migration under an applied electric field was simulated according to the force in Eq.~\ref{eq:total-force}. The interval for charge evaluation using OFDFT was set to 250~fs. In this structure, $\beta$-\LiPS\ was arranged in a $2\times2$ array in the plane, cleaved along the (001) surface, and six $\mathrm{S_8}$ rings were placed between the slabs; this arrangement was periodically repeated. The structure was relaxed by Langevin dynamics at 300~K for 10~ps, followed by Langevin dynamics at 100~K for 10~ps to remove impulsive forces between atoms. Figures~\ref{fig:heterto_short}(b) and \ref{fig:heterto_short}(c) plot the time evolution of the $z$ coordinates of representative Li ions when an electric field was applied at a temperature of 500~K. As shown in Fig.~\ref{fig:heterto_short}(b), when the electric field was small, the Li ions did not overcome the interfacial energy barrier within a simulation time of approximately 170~ps; when the electric field was stronger, the Li ions were observed to diffuse across the layers, as shown in Fig.~\ref{fig:heterto_short}(c).

\begin{figure}[ht]
    \centering
    \includegraphics[width=0.37 \textwidth]{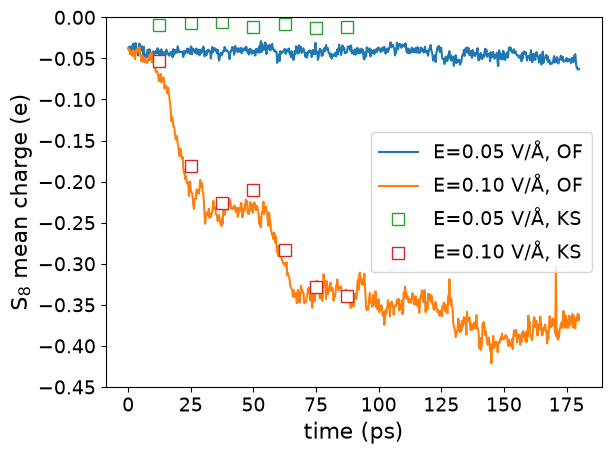}
    \caption{
Mean charge of the 48 S atoms that formed the $\mathrm{S_8}$ rings in the initial structure ($T=500$~K). The atoms are initially neutral but become negatively charged on average as Li ions enter from the \LiPS\ region during the time evolution when an electric field of $0.10~\mathrm{V\AA^{-1}}$ is applied.
Square markers denote the charge evaluation by KS-PAW for the snapshot structures in the OFDFT-assisted MD trajectory.
    }
    \label{fig:s_chg_short}
\end{figure}

The change in the charges of the S atoms accompanying this Li migration is shown in Fig.~\ref{fig:s_chg_short}. The figure shows the evolution of the mean charge of the S atoms that formed the neutral $\mathrm{S_8}$-ring layer in the initial structure. As Li ions entered the layer, the OFDFT electron-density optimization calculation for the entire simulation cell captured the negative charging of the S atoms.

To evaluate the quantitative accuracy of the OFDFT charges during the MD simulation, KSDFT calculations using the PAW method and the PZ functional were performed for the snapshot structures. 
Fig.~\ref{fig:s_chg_short} shows that the reduction of the S atoms is also supported by Bader charge analysis based on KS-DFT.
In Fig.~\ref{fig:heterto_short}(c), and the resulting charges of the structure at 50,000~fs were compared.
As shown in Table~\ref{tab:compare_chg}, the OFDFT and KSDFT Li charges are in good agreement, as in the bulk case. Although the S charges differ by approximately 0.1, it was confirmed at the KSDFT level of accuracy that the negative charging of S caused by the arrival of Li is not a behavior found only in OFDFT.

Because the present calculation examines transient behavior before equilibrium is reached, it is difficult to extract physically meaningful statistical quantities.
However, when charges change owing to changes in the local environment as described above, they cannot be tracked by MD using a classical force field with initial charges assigned. Incorporating charge evaluation using OFDFT is therefore considered one possible solution for simulations of electric-field-driven ionic systems.
\begin{figure}[ht]
    \centering
    \includegraphics[width=0.32 \textwidth]{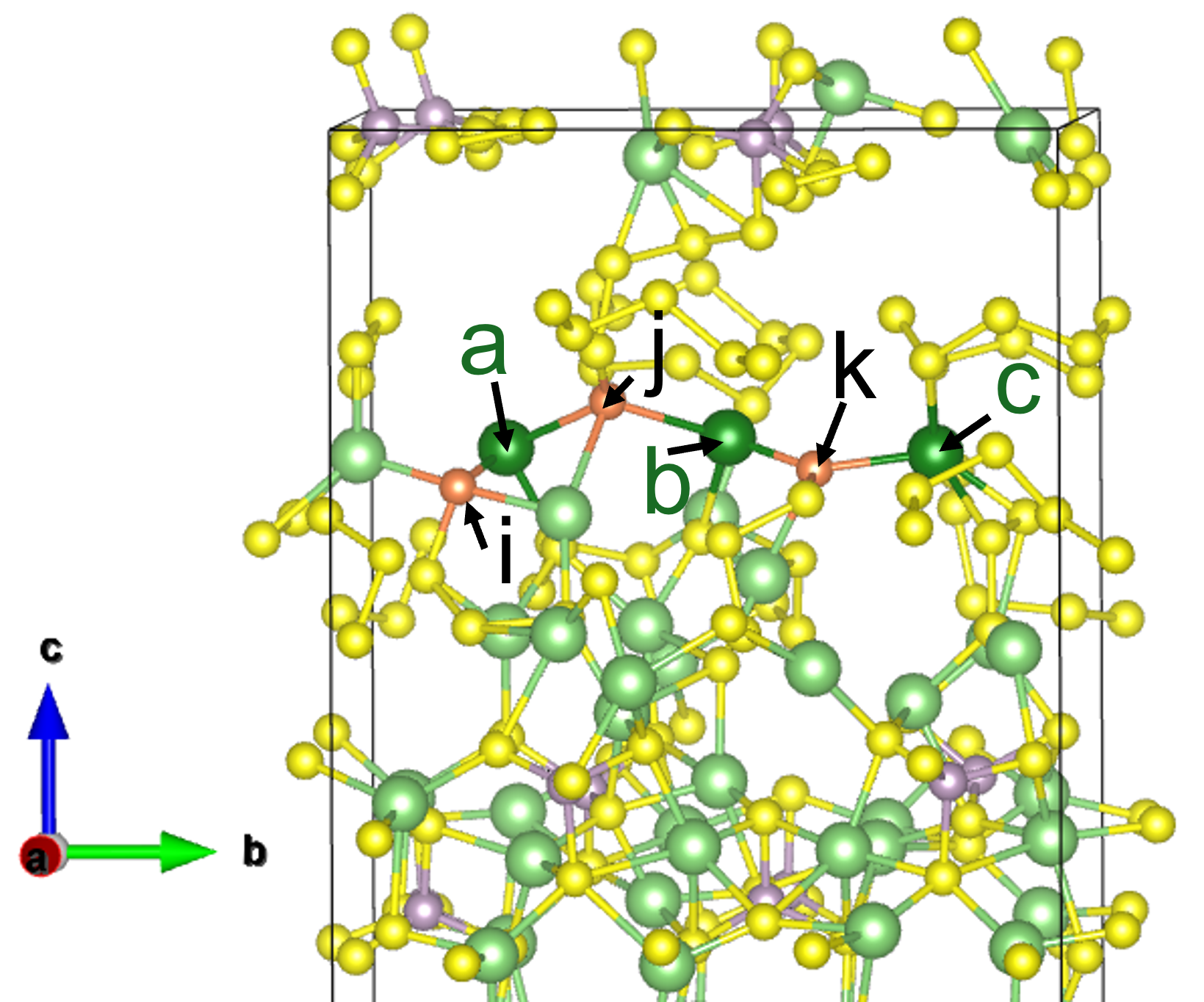}
    \caption{
Snapshot of the interfacial structure at $t=50,000$~fs in Fig.~\ref{fig:heterto_short}(c). The charges were compared for the representative Li atoms shown in dark green and S atoms shown in orange.
    }
    \label{fig:interface_t50000}
\end{figure}
\begin{table}[ht]
\centering
\caption{Comparison of KSDFT and OFDFT charges}
\label{tab:compare_chg}
\begin{tabular}{lcc} \hline
   Atom & KSDFT & OFDFT \\ \hline
   Li, a  & +0.89 & +0.86  \\
   Li, b  & +0.88 & +0.86  \\
   Li, c & +0.87 & +0.87 \\
   S, i  & -0.73 & -0.60  \\
   S, j  & -0.67 & -0.69  \\
   S, k & -0.76 & -0.63 \\ \hline
\end{tabular}
\end{table}

\section{Conclusions}
In this study, we developed an OFDFT-assisted machine-learned molecular dynamics framework for ionic systems.
While utilizing the efficiency and force accuracy of a modern machine-learned interatomic potential for field-independent interatomic forces, atomic charges are evaluated at fixed intervals based on electron-density calculations using OFDFT.
Application to a model electrode--electrolyte interface in a Li battery material enabled us to observe the migration of Li ions along the electric field and changes in the charges of S atoms in the electrode depending on their local environments. There is no need to call computationally expensive KSDFT for charge evaluation, and because a pretrained general-purpose machine-learned potential was used in the present demonstration, there is also no need to prepare a machine-learning model for charge prediction or a vast amount of KSDFT-based training data for fine-tuning the forces. DFT-based charge evaluation is also possible for elements such as S that can adopt various valence states. Future work includes verifying the generality of the method by applying it to various systems and identifying use cases in which information useful for actual materials development and performance evaluation can be extracted by reproducing experimental observations.


\begin{acknowledgments}
This work was partially supported by the Center of Innovations for Sustainable Quantum AI (JST Grant Number
JPMJPF2221). The computation in this work has been done using the facilities of the Supercomputer Center, the
Institute for Solid State Physics, the University of Tokyo (ISSPkyodo-SC-2026-Ea-0014).

\end{acknowledgments}



\bibliographystyle{apsrev4-2}
\bibliography{references}

@article{Islam2015,
  author  = {Islam, Md Mahbubul and Ostadhossein, Alireza and Borodin, Oleg and Yeates, A. Todd and Tipton, William W. and Hennig, Richard G. and Kumar, Nitin and van Duin, Adri C. T.},
  title   = {{ReaxFF} Molecular Dynamics Simulations on Lithiated Sulfur Cathode Materials},
  journal = {Physical Chemistry Chemical Physics},
  year    = {2015},
  volume  = {17},
  number  = {5},
  pages   = {3383--3393},
  doi     = {10.1039/C4CP04532G},
  url     = {https://doi.org/10.1039/C4CP04532G}
}

@article{Senftle2016,
  author  = {Senftle, Thomas P. and Hong, Sungwook and Islam, Md Mahbubul and Kylasa, Sudhir B. and Zheng, Yuanxia and Shin, Yun Kyung and Junkermeier, Chad and Engel-Herbert, Roman and Janik, Michael J. and Aktulga, Hasan Metin and Verstraelen, Toon and Grama, Ananth and van Duin, Adri C. T.},
  title   = {The {ReaxFF} Reactive Force-Field: Development, Applications and Future Directions},
  journal = {npj Computational Materials},
  year    = {2016},
  volume  = {2},
  pages   = {15011},
  doi     = {10.1038/npjcompumats.2015.11},
  url     = {https://doi.org/10.1038/npjcompumats.2015.11}
}

@article{Shimizu2023,
author = {Koji Shimizu and Ryuji Otsuka and Masahiro Hara and Emi Minamitani and Satoshi Watanabe},
title = {Prediction of Born effective charges using neural network to study ion migration under electric fields: applications to crystalline and amorphous Li3PO4},
journal = {Science and Technology of Advanced Materials: Methods},
volume = {3},
number = {1},
pages = {2253135},
year = {2023},
publisher = {Taylor \& Francis},
doi = {10.1080/27660400.2023.2253135},


URL = { 
    
        https://doi.org/10.1080/27660400.2023.2253135
    
    

},
eprint = { 
    
        https://doi.org/10.1080/27660400.2023.2253135
    
    

}

}

@article{Batatia2025,
  author  = {Batatia, Ilyes and Benner, Philipp and Chiang, Yuan and Elena, Alin M. and Kov{\'a}cs, D{\'a}vid P. and Riebesell, Janosh and others},
  title   = {A Foundation Model for Atomistic Materials Chemistry},
  journal = {The Journal of Chemical Physics},
  year    = {2025},
  volume  = {163},
  number  = {18},
  pages   = {184110},
  doi     = {10.1063/5.0297006},
  url     = {https://doi.org/10.1063/5.0297006}
}

@article{Rhodes2025,
  author        = {Rhodes, Benjamin and Vandenhaute, Sander and Simkus, Vaidotas and Gin, James and Godwin, Jonathan and Duignan, Tim and Neumann, Mark},
  title         = {{Orb-v3}: Atomistic Simulation at Scale},
  journal       = {arXiv preprint arXiv:2504.06231},
  year          = {2025},
  eprint        = {2504.06231},
  archivePrefix = {arXiv},
  primaryClass  = {cond-mat.mtrl-sci},
  doi           = {10.48550/arXiv.2504.06231},
  url           = {https://doi.org/10.48550/arXiv.2504.06231}
}

@article{Shao2021,
  author  = {Shao, Xuecheng and Jiang, Kaili and Mi, Wenhui and Genova, Alessandro and Pavanello, Michele},
  title   = {{DFTpy}: An Efficient and Object-Oriented Platform for Orbital-Free {DFT} Simulations},
  journal = {WIREs Computational Molecular Science},
  year    = {2021},
  volume  = {11},
  number  = {1},
  pages   = {e1482},
  doi     = {10.1002/wcms.1482},
  url     = {https://doi.org/10.1002/wcms.1482}
}

@article{Shao_revHC2021,
  title = {Revised Huang-Carter nonlocal kinetic energy functional for semiconductors and their surfaces},
  author = {Shao, Xuecheng and Mi, Wenhui and Pavanello, Michele},
  journal = {Phys. Rev. B},
  volume = {104},
  issue = {4},
  pages = {045118},
  numpages = {8},
  year = {2021},
  month = {Jul},
  publisher = {American Physical Society},
  doi = {10.1103/PhysRevB.104.045118},
  url = {https://link.aps.org/doi/10.1103/PhysRevB.104.045118}
}

@article{Gonzalez2002,
  title = {Dynamical properties of liquid Al near melting:  An orbital-free molecular dynamics study},
  author = {Gonz\'alez, D. J. and Gonz\'alez, L. E. and L\'opez, J. M. and Stott, M. J.},
  journal = {Phys. Rev. B},
  volume = {65},
  issue = {18},
  pages = {184201},
  numpages = {13},
  year = {2002},
  month = {Apr},
  publisher = {American Physical Society},
  doi = {10.1103/PhysRevB.65.184201},
  url = {https://link.aps.org/doi/10.1103/PhysRevB.65.184201}
}

@article{Luo2018,
  author  = {Luo, Kai and Karasiev, Valentin V. and Trickey, S. B.},
  title   = {A Simple Generalized Gradient Approximation for the Noninteracting Kinetic Energy Density Functional},
  journal = {Physical Review B},
  year    = {2018},
  volume  = {98},
  pages   = {041111},
  doi     = {10.1103/PhysRevB.98.041111},
  url     = {https://doi.org/10.1103/PhysRevB.98.041111}
}

@article{PerdewWang1992,
  author  = {Perdew, John P. and Wang, Yue},
  title   = {Accurate and simple analytic representation of the
             electron-gas correlation energy},
  journal = {Phys. Rev. B},
  volume  = {45},
  number  = {23},
  pages   = {13244--13249},
  year    = {1992},
  doi     = {10.1103/PhysRevB.45.13244}
}

@article{PerdewZunger1981,
  author  = {Perdew, John P. and Zunger, Alex},
  title   = {Self-interaction correction to density-functional
             approximations for many-electron systems},
  journal = {Phys. Rev. B},
  volume  = {23},
  number  = {10},
  pages   = {5048--5079},
  year    = {1981},
  doi     = {10.1103/PhysRevB.23.5048}
}

@article{Henkelman2006,
  author  = {Henkelman, Graeme and Arnaldsson, Andri and Jonsson, Hannes},
  title   = {A Fast and Robust Algorithm for {Bader} Decomposition of Charge Density},
  journal = {Computational Materials Science},
  year    = {2006},
  volume  = {36},
  number  = {3},
  pages   = {354--360},
  doi     = {10.1016/j.commatsci.2005.04.010},
  url     = {https://doi.org/10.1016/j.commatsci.2005.04.010}
}

@article{Giannozzi2009,
  author  = {Giannozzi, Paolo and Baroni, Stefano and Bonini, Nicola and Calandra, Matteo and Car, Roberto and Cavazzoni, Carlo and Ceresoli, Davide and Chiarotti, Guido L. and Cococcioni, Matteo and Dabo, Ismaila and Dal Corso, Andrea and de Gironcoli, Stefano and Fabris, Stefano and Fratesi, Guido and Gebauer, Ralph and Gerstmann, Uwe and Gougoussis, Christos and Kokalj, Anton and Lazzeri, Michele and Martin-Samos, Layla and Marzari, Nicola and Mauri, Francesco and Mazzarello, Riccardo and Paolini, Stefano and Pasquarello, Alfredo and Paulatto, Lorenzo and Sbraccia, Carlo and Scandolo, Sandro and Sclauzero, Gabriele and Seitsonen, Ari P. and Smogunov, Alexander and Umari, Paolo and Wentzcovitch, Renata M.},
  title   = {{QUANTUM ESPRESSO}: A Modular and Open-Source Software Project for Quantum Simulations of Materials},
  journal = {Journal of Physics: Condensed Matter},
  year    = {2009},
  volume  = {21},
  number  = {39},
  pages   = {395502},
  doi     = {10.1088/0953-8984/21/39/395502},
  url     = {https://doi.org/10.1088/0953-8984/21/39/395502}
}

@article{Giannozzi2017,
  author  = {Giannozzi, Paolo and Andreussi, Oliviero and Brumme, Till and Bunau, Oana and Buongiorno Nardelli, Marco and Calandra, Matteo and Car, Roberto and Cavazzoni, Carlo and Ceresoli, Davide and Cococcioni, Matteo and Colonna, Nicola and Carnimeo, Ivan and Dal Corso, Andrea and de Gironcoli, Stefano and Delugas, Pietro and DiStasio, Robert A. and Ferretti, Andrea and Floris, Andrea and Fratesi, Guido and Fugallo, Giorgia and Gebauer, Ralph and Gerstmann, Uwe and Giustino, Feliciano and Gorni, Tiziano and Jia, Ji and Kawamura, Mitsuaki and Ko, Hikaru-Y. and Kokalj, Anton and Kucukbenli, Emine and Lazzeri, Michele and Marsili, Margherita and Marzari, Nicola and Mauri, Francesco and Nguyen, Nguyen L. and Nguyen, Huy-Viet and Otero-de-la-Roza, Alberto and Paulatto, Lorenzo and Ponce, Samuel and Rocca, Dario and Sabatini, Riccardo and Santra, Biswajit and Schlipf, Martin and Seitsonen, Ari P. and Smogunov, Alexander and Timrov, Iurii and Thonhauser, Timo and Umari, Paolo and Vast, Nathalie and Wu, Xifan and Baroni, Stefano},
  title   = {Advanced Capabilities for Materials Modelling with {Quantum ESPRESSO}},
  journal = {Journal of Physics: Condensed Matter},
  year    = {2017},
  volume  = {29},
  number  = {46},
  pages   = {465901},
  doi     = {10.1088/1361-648X/aa8f79},
  url     = {https://doi.org/10.1088/1361-648X/aa8f79}
}

@article{Larsen2017,
  author  = {Larsen, Ask Hjorth and Mortensen, Jens Jorgen and Blomqvist, Jakob and Castelli, Ivano E. and Christensen, Rune and Dulak, Marcin and Friis, Jesper and Groves, Michael N. and Hammer, Bjork and Hargus, Cory and Hermes, Eric D. and Jennings, Paul C. and Jensen, Peter Bjerre and Kermode, James and Kitchin, John R. and Kolsbjerg, Esben L. and Kubal, Joseph and Kaasbjerg, Kristen and Lysgaard, Steen and Maronsson, Jon Bergmann and Maxson, Tristan and Olsen, Thomas and Pastewka, Lars and Peterson, Andrew and Rostgaard, Carsten and Schiotz, Jakob and Schutt, Ole and Strange, Mikkel and Thygesen, Kristian S. and Vegge, Tejs and Vilhelmsen, Lasse and Walter, Michael and Zeng, Zhenhua and Jacobsen, Karsten W.},
  title   = {The Atomic Simulation Environment - A Python Library for Working with Atoms},
  journal = {Journal of Physics: Condensed Matter},
  year    = {2017},
  volume  = {29},
  number  = {27},
  pages   = {273002},
  doi     = {10.1088/1361-648X/aa680e},
  url     = {https://doi.org/10.1088/1361-648X/aa680e}
}

@article{Nose1984,
  author  = {Nose, Shuichi},
  title   = {A Unified Formulation of the Constant Temperature Molecular Dynamics Methods},
  journal = {Journal of Chemical Physics},
  year    = {1984},
  volume  = {81},
  number  = {1},
  pages   = {511--519},
  doi     = {10.1063/1.447334},
  url     = {https://doi.org/10.1063/1.447334}
}

@article{Hoover1985,
  author  = {Hoover, William G.},
  title   = {Canonical Dynamics: Equilibrium Phase-Space Distributions},
  journal = {Physical Review A},
  year    = {1985},
  volume  = {31},
  number  = {3},
  pages   = {1695--1697},
  doi     = {10.1103/PhysRevA.31.1695},
  url     = {https://doi.org/10.1103/PhysRevA.31.1695}
}

@article{HohenbergKohn1964,
  title = {Inhomogeneous Electron Gas},
  author = {Hohenberg, P. and Kohn, W.},
  journal = {Phys. Rev.},
  volume = {136},
  issue = {3B},
  pages = {B864--B871},
  numpages = {0},
  year = {1964},
  month = {Nov},
  publisher = {American Physical Society},
  doi = {10.1103/PhysRev.136.B864},
  url = {https://doi.org/10.1103/PhysRev.136.B864}
}

@article{KohnSham1965,
  title = {Self-Consistent Equations Including Exchange and Correlation Effects},
  author = {Kohn, W. and Sham, L. J.},
  journal = {Phys. Rev.},
  volume = {140},
  issue = {4A},
  pages = {A1133--A1138},
  numpages = {0},
  year = {1965},
  month = {Nov},
  publisher = {American Physical Society},
  doi = {10.1103/PhysRev.140.A1133},
  url = {https://doi.org/10.1103/PhysRev.140.A1133}
}

@article{PAW1994,
  title = {Projector augmented-wave method},
  author = {Bl\"ochl, P. E.},
  journal = {Phys. Rev. B},
  volume = {50},
  issue = {24},
  pages = {17953--17979},
  numpages = {0},
  year = {1994},
  month = {Dec},
  publisher = {American Physical Society},
  doi = {10.1103/PhysRevB.50.17953},
  url = {https://doi.org/10.1103/PhysRevB.50.17953}
}

@article{Imoto2021PRR,
  title = {Order-$N$ orbital-free density-functional calculations with machine learning of functional derivatives for semiconductors and metals},
  author = {Imoto, Fumihiro and Imada, Masatoshi and Oshiyama, Atsushi},
  journal = {Phys. Rev. Res.},
  volume = {3},
  issue = {3},
  pages = {033198},
  numpages = {17},
  year = {2021},
  month = {Aug},
  publisher = {American Physical Society},
  doi = {10.1103/PhysRevResearch.3.033198},
  url = {https://doi.org/10.1103/PhysRevResearch.3.033198}
}

@article{delRioCarter_review2018,
  title   = {Orbital-Free Density Functional Theory for Materials Research},
  author  = {Witt, William C. and del Rio, Beatriz G. and Dieterich, Johannes M. and Carter, Emily A.},
  journal = {Journal of Materials Research},
  year    = {2018},
  volume  = {33},
  number  = {7},
  pages   = {777--795},
  doi     = {10.1557/jmr.2017.462},
  url     = {https://doi.org/10.1557/jmr.2017.462}
}

@article{Mi_review2023,
  author  = {Mi, Wenhui and Luo, Kai and Trickey, S. B. and Pavanello, Michele},
  title   = {Orbital-Free Density Functional Theory: An Attractive Electronic Structure Method for Large-Scale First-Principles Simulations},
  journal = {Chemical Reviews},
  year    = {2023},
  volume  = {123},
  number  = {21},
  pages   = {12039--12104},
  doi     = {10.1021/acs.chemrev.2c00758},
  url     = {https://doi.org/10.1021/acs.chemrev.2c00758}
}

@article{HuangCarter2010PRB,
  title = {Nonlocal orbital-free kinetic energy density functional for semiconductors},
  author = {Huang, Chen and Carter, Emily A.},
  journal = {Phys. Rev. B},
  volume = {81},
  issue = {4},
  pages = {045206},
  numpages = {15},
  year = {2010},
  month = {Jan},
  publisher = {American Physical Society},
  doi = {10.1103/PhysRevB.81.045206},
  url = {https://doi.org/10.1103/PhysRevB.81.045206}
}

@article{Zhou2004PRB,
  title = {Transferable local pseudopotentials derived via inversion of the Kohn-Sham equations in a bulk environment},
  author = {Zhou, Baojing and Alexander Wang, Yan and Carter, Emily A.},
  journal = {Phys. Rev. B},
  volume = {69},
  issue = {12},
  pages = {125109},
  numpages = {15},
  year = {2004},
  month = {Mar},
  publisher = {American Physical Society},
  doi = {10.1103/PhysRevB.69.125109},
  url = {https://doi.org/10.1103/PhysRevB.69.125109}
}

@Article{Huang2008,
author ="Huang, Chen and Carter, Emily A.",
title  ="Transferable local pseudopotentials for magnesium{,} aluminum and silicon",
journal  ="Phys. Chem. Chem. Phys.",
year  ="2008",
volume  ="10",
issue  ="47",
pages  ="7109-7120",
publisher  ="The Royal Society of Chemistry",
doi  ="10.1039/B810407G",
url  ="https://doi.org/10.1039/B810407G"}

@article{delRio2017,
author = {del Rio, Beatriz G. and Dieterich, Johannes M. and Carter, Emily A.},
title = {Globally-Optimized Local Pseudopotentials for (Orbital-Free) Density Functional Theory Simulations of Liquids and Solids},
journal = {Journal of Chemical Theory and Computation},
volume = {13},
number = {8},
pages = {3684-3695},
year = {2017},
doi = {10.1021/acs.jctc.7b00565},
note = {PMID: 28686438},
url = {https://doi.org/10.1021/acs.jctc.7b00565}
}

@article{OEPP2016,
    author = {Mi, Wenhui and Zhang, Shoutao and Wang, Yanchao and Ma, Yanming and Miao, Maosheng},
    title = {First-principle optimal local pseudopotentials construction via optimized effective potential method},
    journal = {The Journal of Chemical Physics},
    volume = {144},
    number = {13},
    pages = {134108},
    year = {2016},
    month = {04},
    issn = {0021-9606},
    doi = {10.1063/1.4944989},
    url = {https://doi.org/10.1063/1.4944989},
    eprint = {https://pubs.aip.org/aip/jcp/article-pdf/doi/10.1063/1.4944989/13696705/134108\_1\_online.pdf},
}

@article{Tang_2009,
doi = {10.1088/0953-8984/21/8/084204},
url = {https://doi.org/10.1088/0953-8984/21/8/084204},
year = {2009},
month = {jan},
publisher = {},
volume = {21},
number = {8},
pages = {084204},
author = {Tang, W and Sanville, E and Henkelman, G},
title = {A grid-based Bader analysis algorithm without lattice bias},
journal = {Journal of Physics: Condensed Matter}
}

@article{Sanville_2007,
author = {Sanville, Edward and Kenny, Steven D. and Smith, Roger and Henkelman, Graeme},
title = {Improved grid-based algorithm for Bader charge allocation},
journal = {Journal of Computational Chemistry},
volume = {28},
number = {5},
pages = {899-908},
doi = {10.1002/jcc.20575},
url = {https://doi.org/10.1002/jcc.20575},
eprint = {https://onlinelibrary.wiley.com/doi/pdf/10.1002/jcc.20575},
year = {2007}
}

\end{document}